\documentclass[aps,prb,preprint,groupedaddress,floatfix,amsmath,amssymb,showpacs]{revtex4-1}

\usepackage{graphicx}
\usepackage{amsmath,amssymb,amsfonts}
\usepackage{textcomp}
\usepackage{hyperref}
\usepackage{gensymb}
\usepackage{verbatim}
\usepackage{amsmath}
\hypersetup{breaklinks=true,colorlinks=true,urlcolor=black}
\usepackage{color}
\usepackage{soul}
\DeclareGraphicsExtensions{.jpg,.pdf,.png,.eps}

\begin{document}

\def \TPT {Ta$_4$Pd$_3$Te$_{16}$}

\title{Pressure induced change in the electronic state of \TPT}	

\author{Na Hyun Jo}
\author{Li Xiang}
\author{Udhara S. Kaluarachchi }
\author{Morgan Masters}
\author{Kathryn Neilson}
\author{Savannah S. Downing}
\author{Paul C. Canfield}
\author{Sergey L. Bud'ko}

\affiliation{Ames Laboratory, US DOE, and Department of Physics and Astronomy, Iowa State University, Ames, Iowa 50011, USA}

\date{\today}

\begin{abstract}

We present measurements of superconducting transition temperature, resistivity, magnetoresistivity and temperature dependence of the upper critical field of \TPT~ under pressures up to 16.4 kbar. All measured properties have an anomaly at $\sim 2 - 4$ kbar pressure range, in particular there is a maximum in $T_c$  and upper critical field, $H_{c2}(0)$, and minimum in low temperature, normal state resistivity. Qualitatively, the data can be explained considering the density of state at the Fermi level as a dominant parameter.
  
\end{abstract}

\pacs{74.70.Xa, 74.62.Fj, 71.45.Lr}

\maketitle 

\section{Introduction}
The continuing search for materials with unconventional superconductivity and / or superconductivity coexisting and competing with other ground states has recently returned to light the layered \TPT~compound with PdTe$_2$ chains. \cite{mar91a} \TPT~ crystallizes in the $I2/m$ monoclinic, space group system. Its structure has Ta - Pd - Te layers, with Pd atoms forming PdTe$_2$ chains along the $b$-axis.  Altogether \TPT~ can be looked at as a layered compound with quasi-one-dimensional characteristics. \cite{mar91a}  Its  band structure \cite{ale97a,sin14a,lee15a} contains a combination of distinct one- two- and three-dimensional features, making this compound electronically an anisotropic three-dimensional material. Of notice is the proposed nesting between one-dimensional sheets of the Fermi surface. \cite{sin14a,lee15a}

Superconductivity in \TPT~ at $\sim 4.6$ K was reported in Ref. \onlinecite{jia14a}. Whereas the bulk nature of superconductivity and moderate ($\leq 6$) anisotropy of the upper critical field (consistent with anisotropic three-dimensional electronic properties of the compound) was confirmed by several groups \cite{pan15a,jia15a,xux15a,zha16a}, the nature of superconductivity is still under debate. Thermal conductivity measurements \cite{pan15a} suggested nodal superconductivity, whereas detailed analysis of the specific heat capacity \cite{jia15a} and scanning tunneling spectroscopy studies \cite{fan15a,duz15a}  described superconductivity in \TPT~ as multi-band, with anisotropic superconducting gaps, and NMR/NQR data \cite{liz16a} characterized it as $s$-wave, nodeless, superconductivity. 

In addition to superconductivity, charge density wave (CDW) formation was suggested in \TPT~. \cite{fan15a,duz15a,liz16a,che15a,hel17a} Based on NMR / NQR \cite{liz16a}  measurements, CDW ordering sets in at $T^* \sim 20$ K, although Raman scattering \cite{che15a} results suggest possible CDW transition or emergence of CDW fluctuations below 140 - 200 K. Recent electrical resistivity and magnetic susceptibility \cite{hel17a} measurements detected anomalies in the 10 K - 20 K temperature range. In the susceptibility measurements, reported features are shallow and require a background subtraction to be exposed. In resistivity, a distinct feature is seen when the current is flowing along the $a^*$-axis, perpendicular to quasi-one-dimensional chains in the crystal structure. \cite{hel17a}  The band structure of \TPT~ \cite{sin14a,lee15a} does contain features consistent with possibility of a CDW formation.

All this makes \TPT~ one of the rather rare materials with potentially competing electronic (superconductivity) and charge orders \cite{gab02a} and as such merits further study. One of the accepted approaches to gain additional information on the systems with competing orders is to study changes caused by controlled external perturbations, like chemical substitution, pressure, and magnetic field. Indeed, initial pressure studies of  \TPT~ up to $\sim 22$ kbar via zero field resistance were reported in Ref. \onlinecite{pan15a}, where a superconducting dome in the temperature - pressure phase diagram was observed. In this work, given the potential difficulties of precisely determining values for bulk superconducting transition temperatures solely with resistivity, we first confirm the non-monotonic pressure dependence of bulk $T_c$ using magnetization measurements in addition to resistivity. Then, since the evolution of the upper critical field under pressure has the potential to give insight to the physics of superconducting materials \cite{tau14a,kal16a}, we study the effects of pressure on the upper critical field. Furthermore, we examine the evolution of the electronic subsystem via measurements of the  normal state resistivity and magnetoresistivity of single crystals of \TPT~ under pressure. 

\section{Experimental details}

\begin{figure}
\begin{center}
\includegraphics[angle=0,width=120mm]{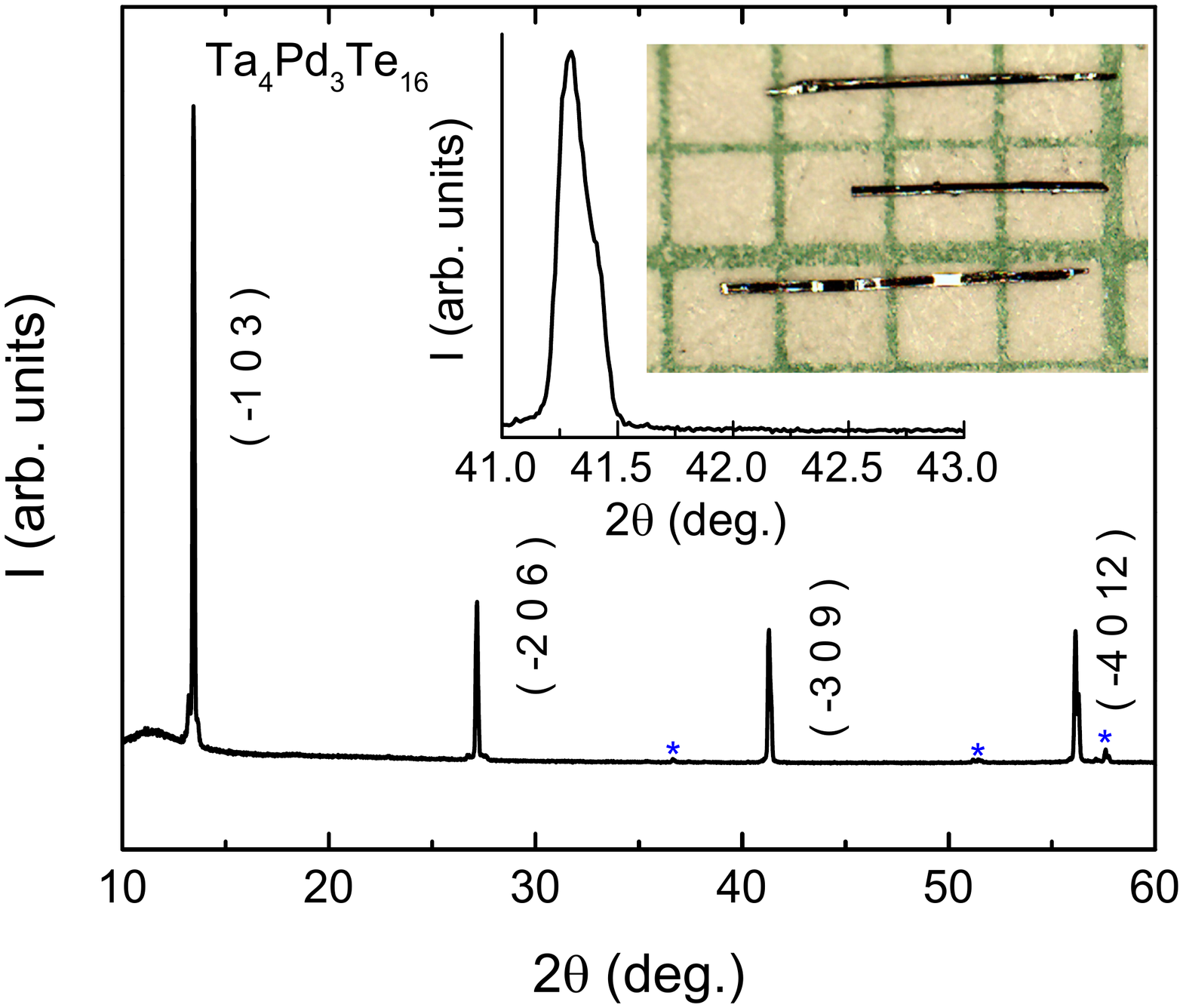}
\end{center}
\caption{(Color online) Single crystal XRD pattern for \TPT~. Blue stars mark Te flux peaks. Part of the diffraction pattern between 41$^{\circ}$ and 43$^{\circ}$, around (-3 0 9) peak, together with typical crystal picture over a mm - scale are shown in the inset.} \label{F1}
\end{figure}

Single crystals of \TPT~were grown by solution method,\cite{can92a} in the way comparable to the reports in Refs. \onlinecite{mar91a,jia14a}. High purity elemental Ta, Pd and Te were placed into an alumina crucible with an alumina frit assemblage \cite{can16a} with the initial stoichiometry of Ta$_{10}$Pd$_{15}$Te$_{75}$, and sealed in an amorphous silica tube.\cite{can92a,can16a} The ampules were heated to 450\,\celsius\, over 3 hours and kept at  450\,\celsius\, for 3 hours, then heated up to 1000\,\celsius\, over 3 hours, kept at 1000\,\celsius\, for 3 hours, cooled down to 700\,\celsius\, over 55 hours, and then finally decanted using a centrifuge.\cite{can92a} The obtained crystals have blade-like morphologies with mirror like surfaces (see an inset to Fig.\ref{F1}). 

A Rigaku MiniFlex II diffractometer (Cu $K_{\alpha}$ radiation) was used for acquiring a single crystal x-ray diffraction (XRD) pattern at room temperature.\cite{jes16a} When the largest surface of the crystal was exposed to x-ray beam, only (-h 0 3l) peaks, where h and l are integers, were detected (Fig.\,\ref{F1}). Small intensity extra peaks marked with blue stars belong to Te flux (seen as silver colored drops on the mirrored faces shown in photo inset to Fig. \ref{F1}). There are no traces of a diffraction peak between 42$^{\circ}$ and 42.5$^{\circ}$ (Fig. \ref{F1}, inset), confirming that the obtained crystals are \TPT, and not the neighboring Ta$_{3}$Pd$_{3}$Te$_{14}$ phase, with similar morphology. \cite{jia16a}

Magnetic measurements under pressure were performed in a Quantum Design, Magnetic Property Measurement System (MPMS), SQUID magnetometer using a commercial HMD piston-cylinder  pressure cell \cite{HMD} with Daphne 7373 oil as a pressure medium (solidifies at $\sim 22$ kbar at room temperature \cite{yok07a}).  Elemental Pb was used as a pressure gauge at low temperatures. \cite{eil81a}

ac electrical transport measurements under pressure were performed using a Quantum Design Physical Property Measurement System (PPMS). A Be-Cu/Ni-Cr-Al hybrid piston-cylinder cell, similar to that used in Ref.\,\onlinecite{bud84a}, was used for pressures up to  $\sim 16.4$ kbar. For this pressure cell, a 40:60 mixture of light mineral oil and $n$-pentane  that solidifies at $\sim30 - 40$ kbar at room temperature \cite{tor15a} was used as a pressure medium. The pressure was determined by the superconducting transition temperature of Pb\cite{eil81a} measured resistively.  The contacts for the electrical transport measurement were prepared in two steps. Firstly, Au contact pads were sputtered on the sample using a simple mask for a standard four-probe configuration. After that, four Pt wires (25\,$\mu$m diameter) were attached on the Au sputtered spots using Epotek-H20E silver epoxy. The contact resistance values were all less than 1 $\ohm$. For these measurements the current was flowing along the $b$ - direction and the magnetic field was applied perpendicular to the mirror-like surface of the sample, along [-1~0~3] direction.

\section{Results}  

\subsection{Superconducting properties}

\begin{figure}
\begin{center}
\includegraphics[angle=0,width=120mm]{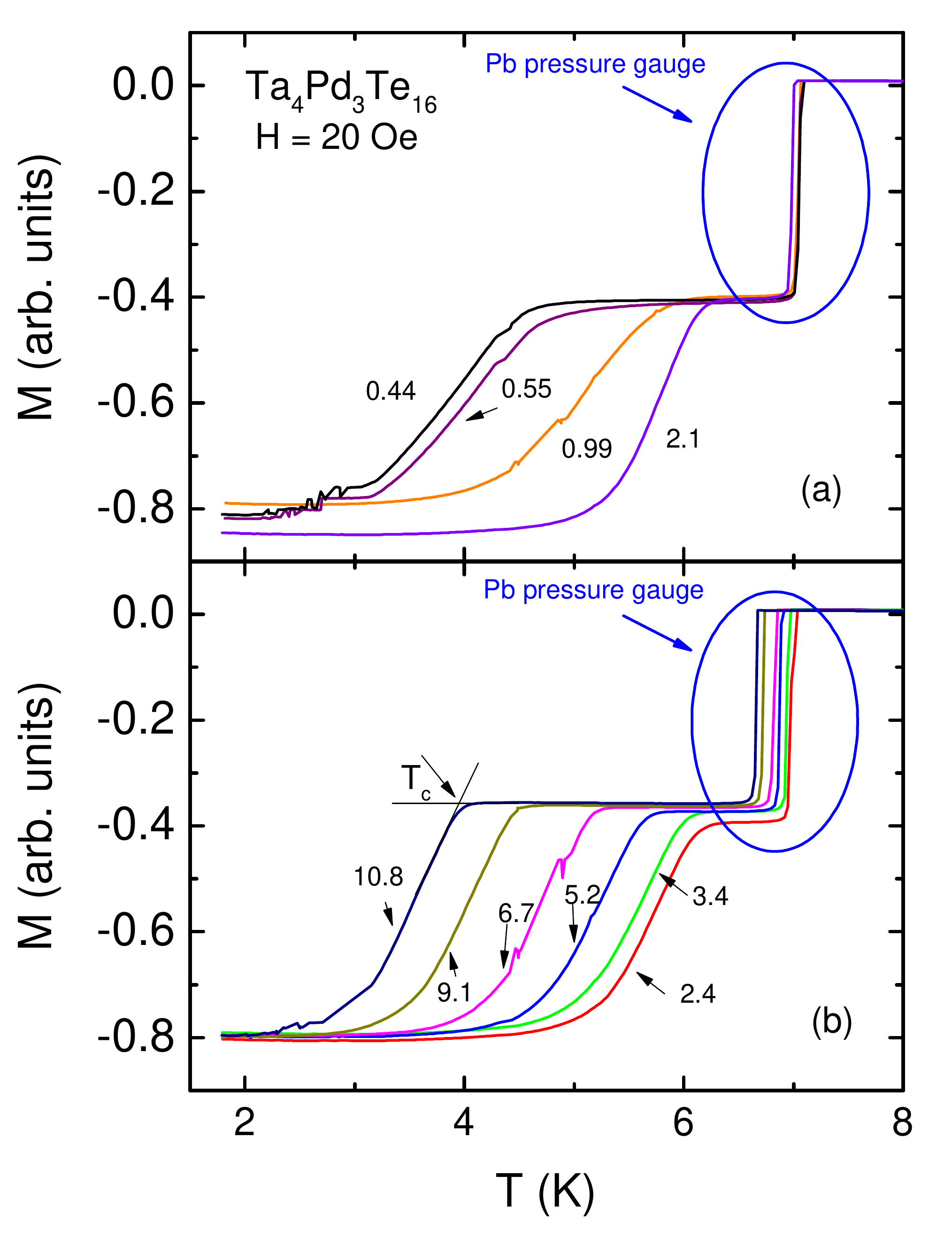}
\end{center}
\caption{(Color online) A subset of zero-field-cooled temperature-dependent magnetization data measured at different pressures (values in kbar are given in the plot). Each curve contains signal from the \TPT~ sample and the Pb pressure gauge. Onset criterion for $T_c$ is shown for the 10.8 kbar data. Panel (a) - lower pressure range with $T_c$ increasing with increasing pressure; panel (b) - higher pressure range with $T_c$ decreasing with increasing pressure.} \label{F2}
\end{figure}

A subset of zero-field-cooled magnetization data measured at different pressures up to $\sim 10.8$ kbar is shown in Fig. \ref{F2}. An onset criterion (shown for the 10.8 kbar curve) was used to determine the $T_c$ values. It is clear that the $T_c(P)$ dependence has a maximum between 2.1 and 2.4 kbar.

\begin{figure}
\begin{center}
\includegraphics[angle=0,width=120mm]{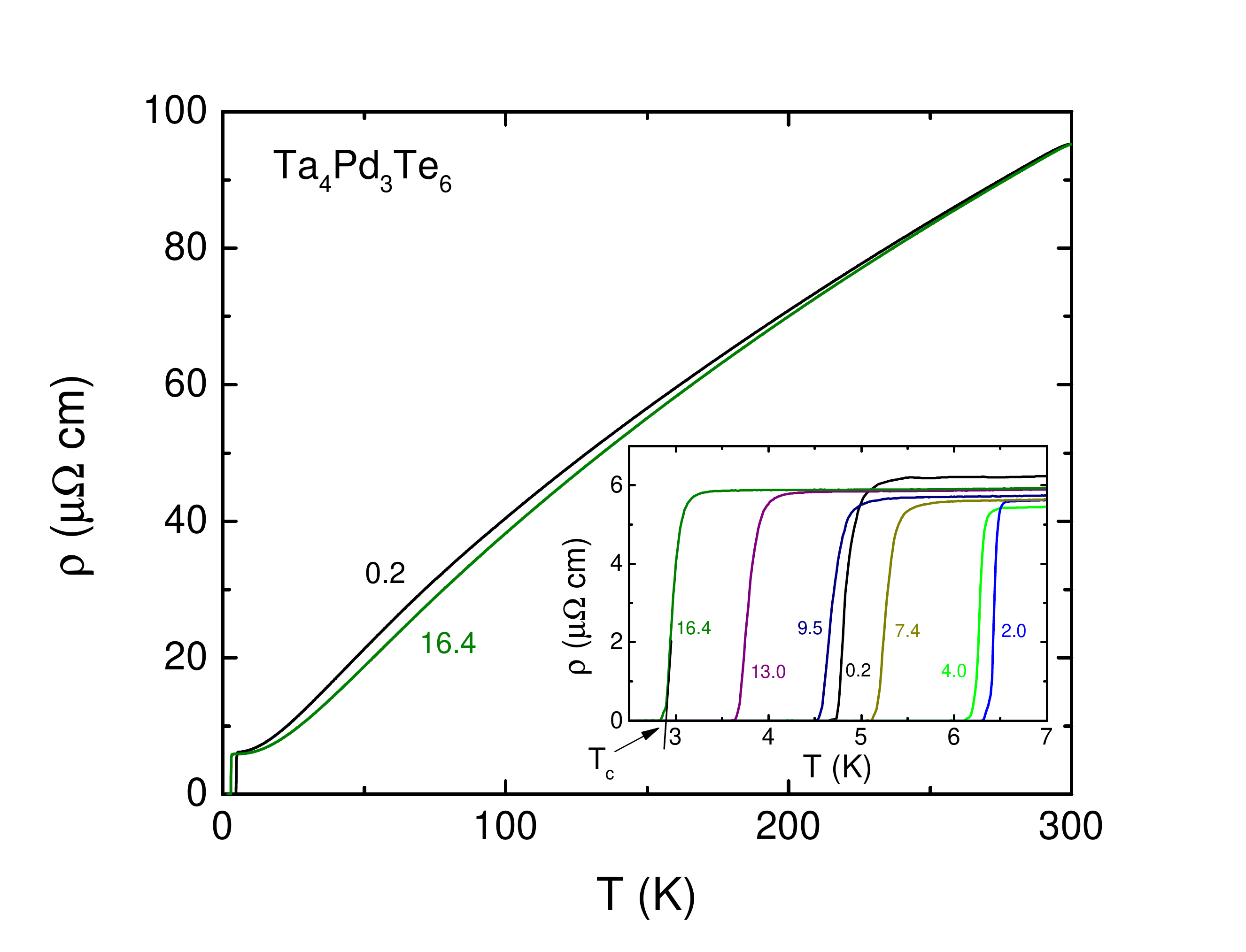}
\end{center}
\caption{(Color online) Zero field resistivity of \TPT~ at $\sim 0.2$ kbar and $\sim 16.4$ kbar (at low temperature) pressure.  Inset: resistive superconducting transitions at several selected pressures (values in kbar are given). Offset criterion for $T_c$ is shown for the 16.4 kbar data in the inset.} \label{F3}
\end{figure}

Zero field resistivity data for the pressure close to ambient ($\sim 0.2$ kbar at low temperature) and for the highest pressure in our measurements  ($\sim 16.4$ kbar at low temperature)  are shown in Fig. \ref{F3}. The residual resistivity ratio, $\rho_{300K}/\rho_{6K} \approx 15.3$, is not far from the values reported at ambient pressure. \cite{jia14a,pan15a} In agreement with the  literature, we do not see any feature than can be associated with CDW in the 0.2 kbar resistivity data measured with the current flowing along the $b$ crystallographic direction. As can be seen in the inset to Fig. \ref{F3}, the $T_c(P)$ behavior is non-monotonic, the low temperature, normal state, resistivity appears to decrease under pressure.

The pressure dependence of the superconducting transition temperature determined from magnetization and resistivity measurements is shown in Fig. \ref{F4}. The two measurement techniques yield very similar results. This suggests that non-monotonic pressure dependence of $T_c$ is a property of the bulk superconducting phase. The overall behavior is consistent with that reported in Ref. \onlinecite{pan15a}. In our data the maximum in $T_c(P)$ is located very close to 2 kbar. At high pressures the pressure derivative is negative and rather large in the absolute value, $dT_c/dP \sim - 0.3$ K/kbar, resulting in 25 - 30 kbar as the extrapolated value of the pressure at which superconductivity will be completely suppressed.

\begin{figure}
\begin{center}
\includegraphics[angle=0,width=120mm]{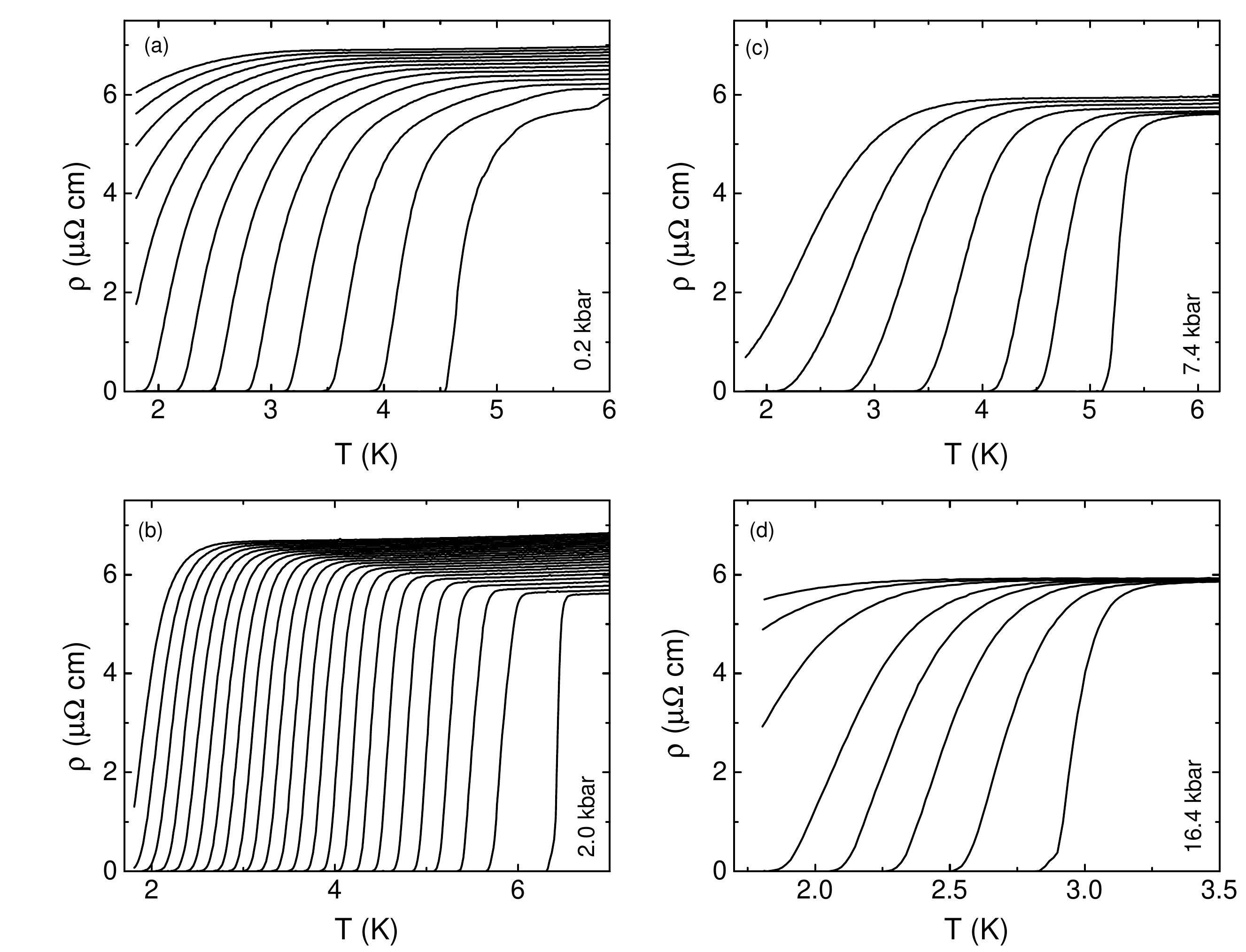}
\end{center}
\caption{Examples of low temperature $\rho(T)$ curves measured at different applied magnetic fields and at different pressures. (a) $P = 0.2$ kbar, $0\leq H \leq 30$ kOe, steps 2.5 kOe; (b)  $P = 2.0$ kbar, $0\leq H \leq 57.5$ kOe, steps 2.5 kOe; (c)  $P =7.4$ kbar, $H = 0, 1.25$ kOe,  $2.5\leq H \leq 12.5$ kOe, steps 2.5 kOe; (d)  $P = 16.4$ kbar, $0\leq H \leq 1$ kOe, steps 0.25 kOe, $1\leq H \leq 2.5$ kOe, steps 0.5 kOe.} \label{F5}
\end{figure}

Examples of the low temperature $\rho(T)$ data measured for different applied magnetic fields and at different pressures are shown in Fig. \ref{F5}.  The data also reveal some positive normal state magnetoresistivity. Based on these data, we were able to compose a manifold of $H_{c2}(T)$ data ($H \| [-1~0~3]$) for different values of pressure (Fig. \ref{F6}(a)). From these data and their derivative, $dH_{c2}(T)/dT$ (Fig. \ref{F6}(b)),  it is clear that there is an upward curvature in $H_{c2}(T)$ at all pressures. The $H_{c2}(T)$ dependencies become close to linear only at about 2 - 3 K. Having in mind the experimentally observed $H_{c2}(T)$ at different pressures, as well as the literature data at ambient pressure down to 0.4 - 1 K, \cite{pan15a,jia15a,zha16a} a reasonable way to evaluate the $H_{c2}(0)$ values appears to be a linear extrapolation to $T =  0$ K of the low temperature part of the curves. 

\begin{figure}
\begin{center}
\includegraphics[angle=0,width=120mm]{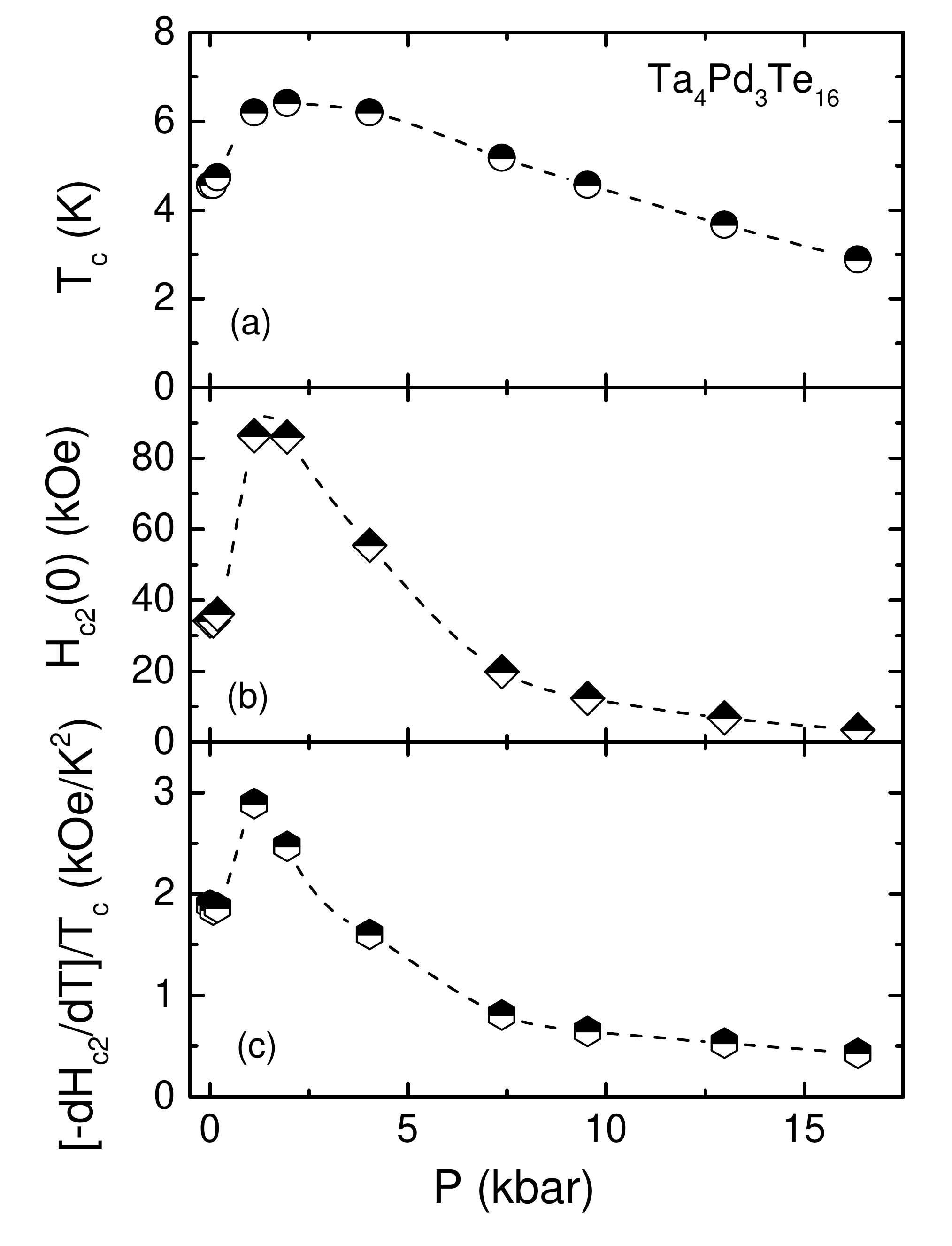}
\end{center}
\caption{Pressure dependences of (a) superconducting transition temperature; (b) extrapolated to $T = 0$  upper critical field (for $H \| (-1~0~3)$); (c) temperature derivatives of the upper critical field (taken at $T = 2$ K) normalized by the $T_c$, $[-dH_{c2}/dT]/T_c$.} \label{F7}
\end{figure}

Pressure dependences of the resistively determined $T_c$, upper critical field extrapolated to $T = 0$ K, $H_{c2}(0)$, and normalized by the respective $T_c$ values temperature derivatives of the upper critical field,   $[-dH_{c2}/dT]/T_c$, are presented in Fig. \ref{F7}. Due to distinct positive curvature of $H_{c2}(T)$ near $T_c$, the values of $dH_{c2}/dT$ for this figure were taken at lower temperatures, where $H_{c2}(T)$ is close to linear. All three superconducting parameters have a maximum at $\sim 2$ kbar, that is somewhat broader in the $T_c(P)$ data, and much more pronounced in $H_{c2}(0)$ vs. $P$ and in $[-dH_{c2}/dT]/T_c$ vs $P$ data. It is noteworthy that after a pressure maximum, both $H_{c2}(0)$ and $[-dH_{c2}/dT]/T_c$ datasets decrease with pressure significantly faster than $T_c(P)$.

\begin{figure}
\begin{center}
\includegraphics[angle=0,width=120mm]{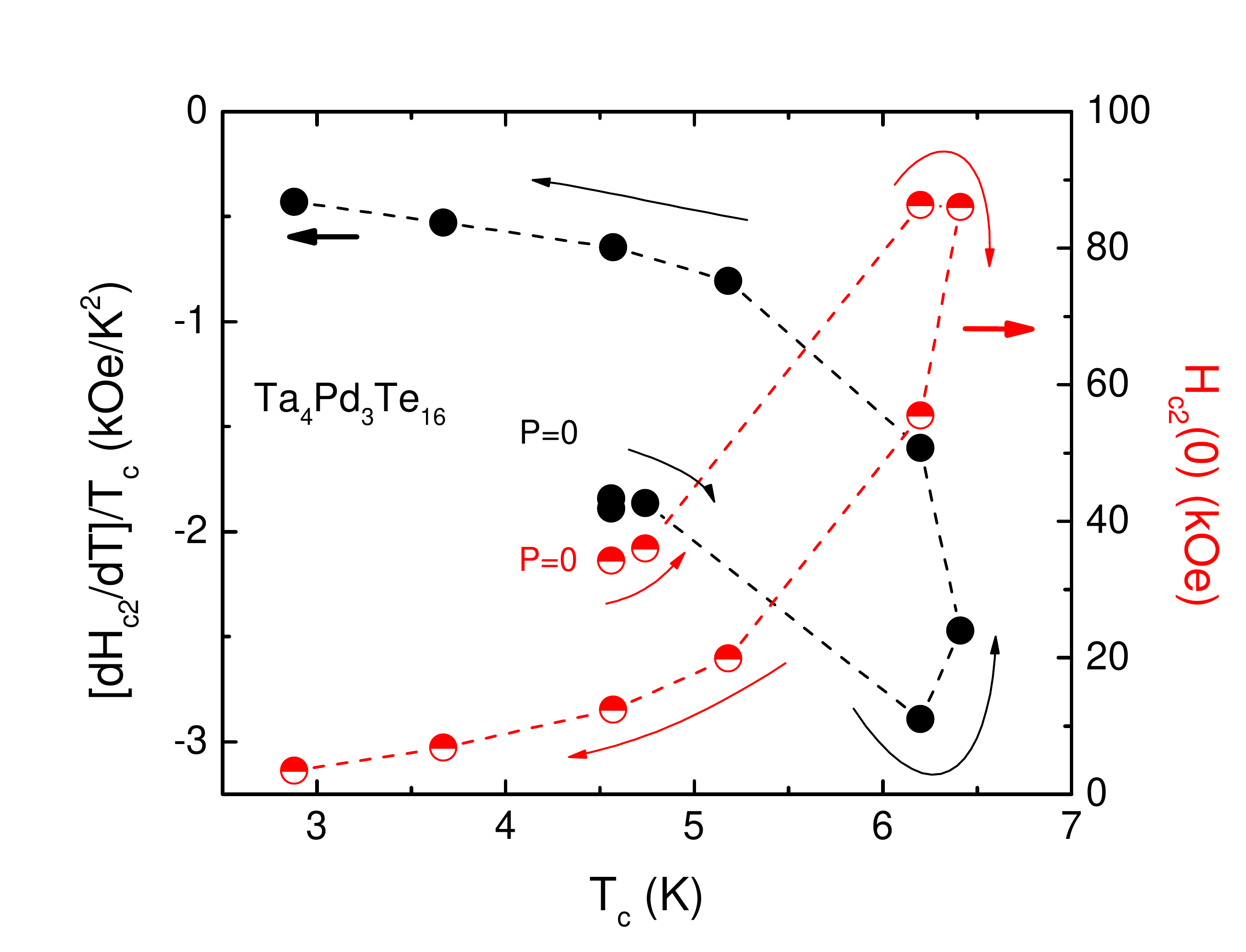}
\end{center}
\caption{(Color online) Normalized temperature derivative of the upper critical field (left Y-axis), and extrapolated to $T = 0$ K upper criticl field, plotted as a function of $T_c$. Arrows indicate the directions of increasing pressure for each manifold. } \label{F8}
\end{figure}

The different pressure dependence of $T_c$ and $[dH_{c2}/dT]/T_c$ is very clear in Fig. \ref{F6}(a) when comparing the $P = 0.2$ kbar and $P = 9.5$ kbar data. To compare low pressure and high pressure superconducting properties more systematically, we plot the $T = 0$ upper critical field and the normalized temperature derivative of  $H_{c2}$ as a function of the superconducting critical temperature (Fig. \ref{F8}).  It is clearly seen that neither of these superconducting parameters scales with $T_c$. Each of the plots has two branches, the low pressure and the high pressure ones,  for the same values of $T_c$, the $H_{c2}(0)$ and the absolute values of $[dH_{c2}/dT]/T_c$ are higher for the lower pressure branch.

\subsection{Normal state properties}

The normal state resistivity for temperatures between 10 K and 300 K,  at different pressures was fitted using the Bloch - Gr\"uneisen - Mott formula that includes interband $s - d$ scattering term: \cite{kac00a}

\begin{equation*}
\rho(T) = \rho_0 + 4RT \left( \frac{T}{\Theta_R} \right)^4  \int_{0}^{\frac{\Theta_D}{T}} \frac{x^5}{(e^x-1)(1-e^{-x})} dx - KT^3
\end{equation*}

The data for two illustrative pressures, together with the fits, are shown in Fig. \ref{F12}(a). The Debye temperature, $\Theta_R$, values, obtained from the fits increase under pressure with a small anomaly in the $\Theta_R (P)$ behavior at $\sim 2 - 4 $ kbar (Fig. \ref{F12}(b)).

\begin{figure}
\begin{center}
\includegraphics[angle=0,width=120mm]{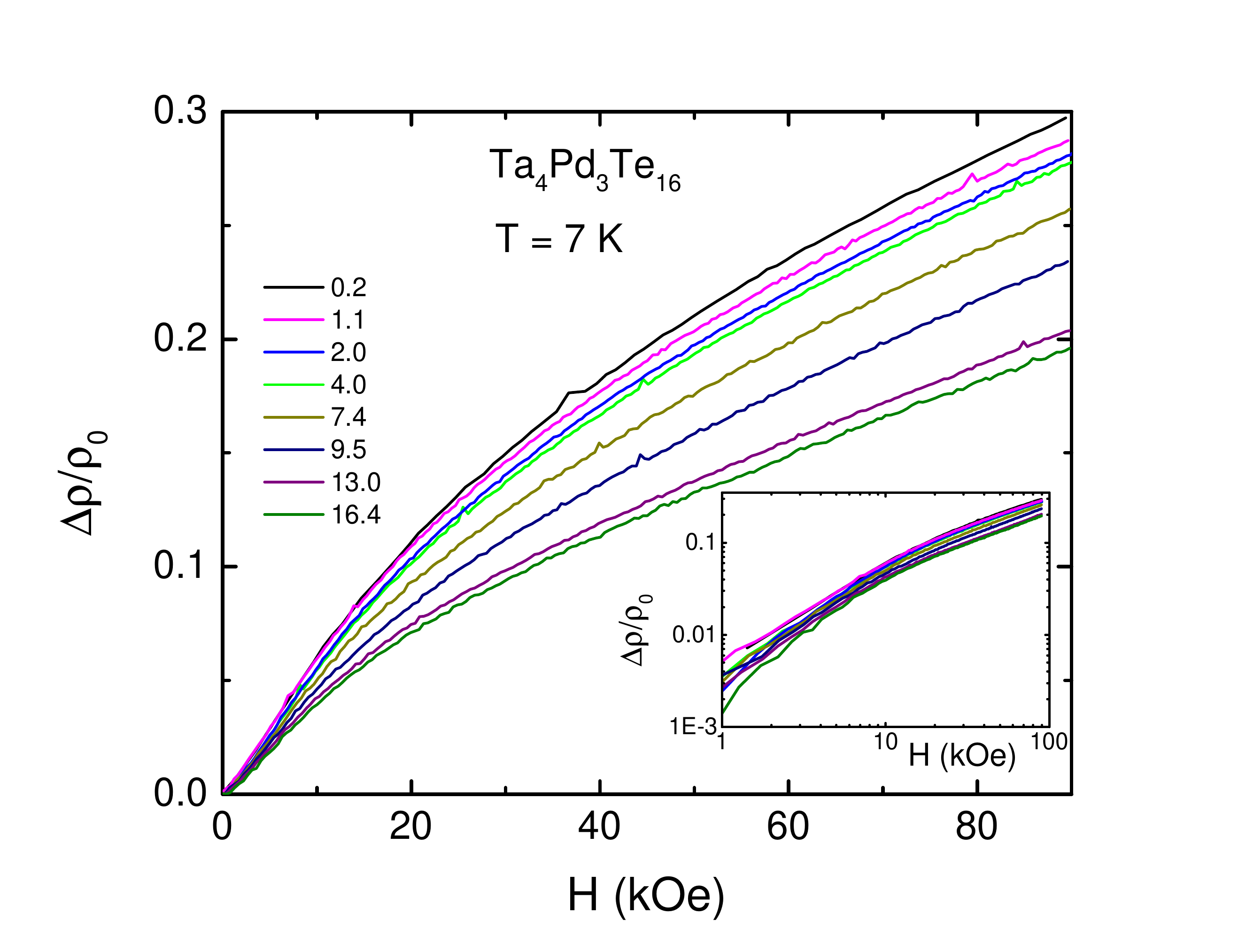}
\end{center}
\caption{(Color online) Normal state, low temperature ($T = 7$ K) magnetoresistivity of \TPT~ plotted as $\Delta \rho / \rho_0 = (\rho_H - \rho_{H = 0})/\rho_{H = 0}$ vs $H$. Inset: the same data on the log-log plot.} \label{F10}
\end{figure}

The magnetic field dependence of low temperature ($T = 7$ K) resistivity is sublinear (Fig. \ref{F10}), in agreement with the ambient pressure report. \cite{xux15a}  The pressure effect on $\Delta \rho / \rho_0 = (\rho_H - \rho_{H = 0})/\rho_{H = 0}$ is significant, although the functional dependence of the  $\Delta \rho / \rho_0 (H)$ appears to be similar at different pressures (Fig. \ref{F10}, inset). The parameters obtained from the magnetoresistivity measurements are plotted in Fig. \ref{F11}. Zero field resistivity at 7 K has a minimum at $P \leq 4$ kbar, and its overall behavior is similar to that of the residual resistivity obtained from low temperature fits to the data (not shown). The resistivity measured at 7 K and 90 kOe initially decreases with pressure, passes through a very shallow minimum, and becomes almost pressure - independent for $P \geq 7.4$ kbar. The pressure dependent magnetoresistivity,  $\Delta\rho_{90}/\rho_0 = (\rho_{H = 90~\text{kOe}} - \rho_{H = 0})/\rho_{H = 0}$, obtained from these data, decreases under pressure with a clear feature in the 2 - 4 kbar pressure range.\\

\section{Discussion and summary}

\begin{figure}
\begin{center}
\includegraphics[angle=0,width=120mm]{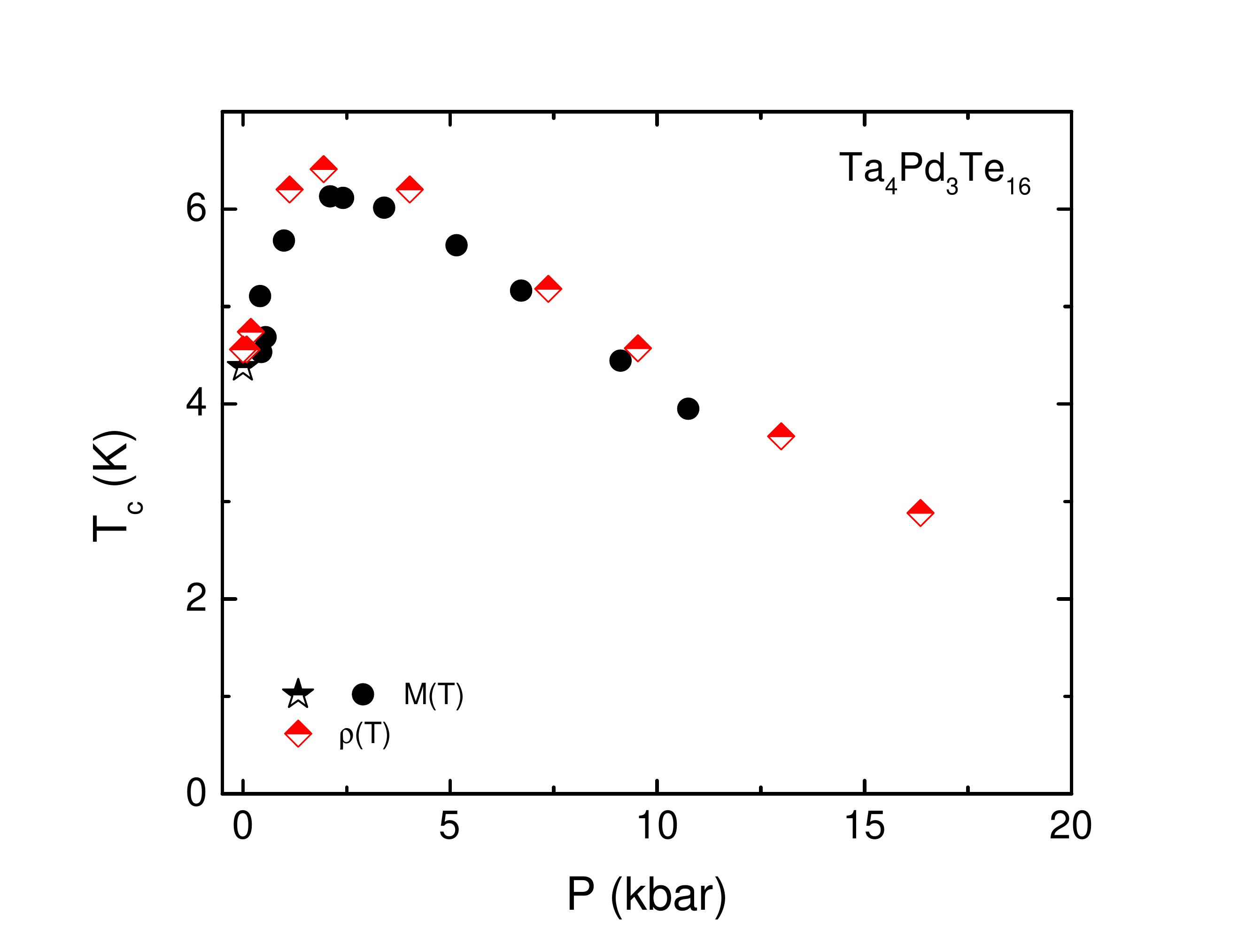}
\end{center}
\caption{(Color online) Pressure dependence of the superconducting transition temperature determined from low field magnetization (circles) and zero field resistivity (rhombi). Star - ambient pressure magnetzation data taken with the sample outside the cell.} \label{F4}
\end{figure}

\begin{figure}
\begin{center}
\includegraphics[angle=0,width=120mm]{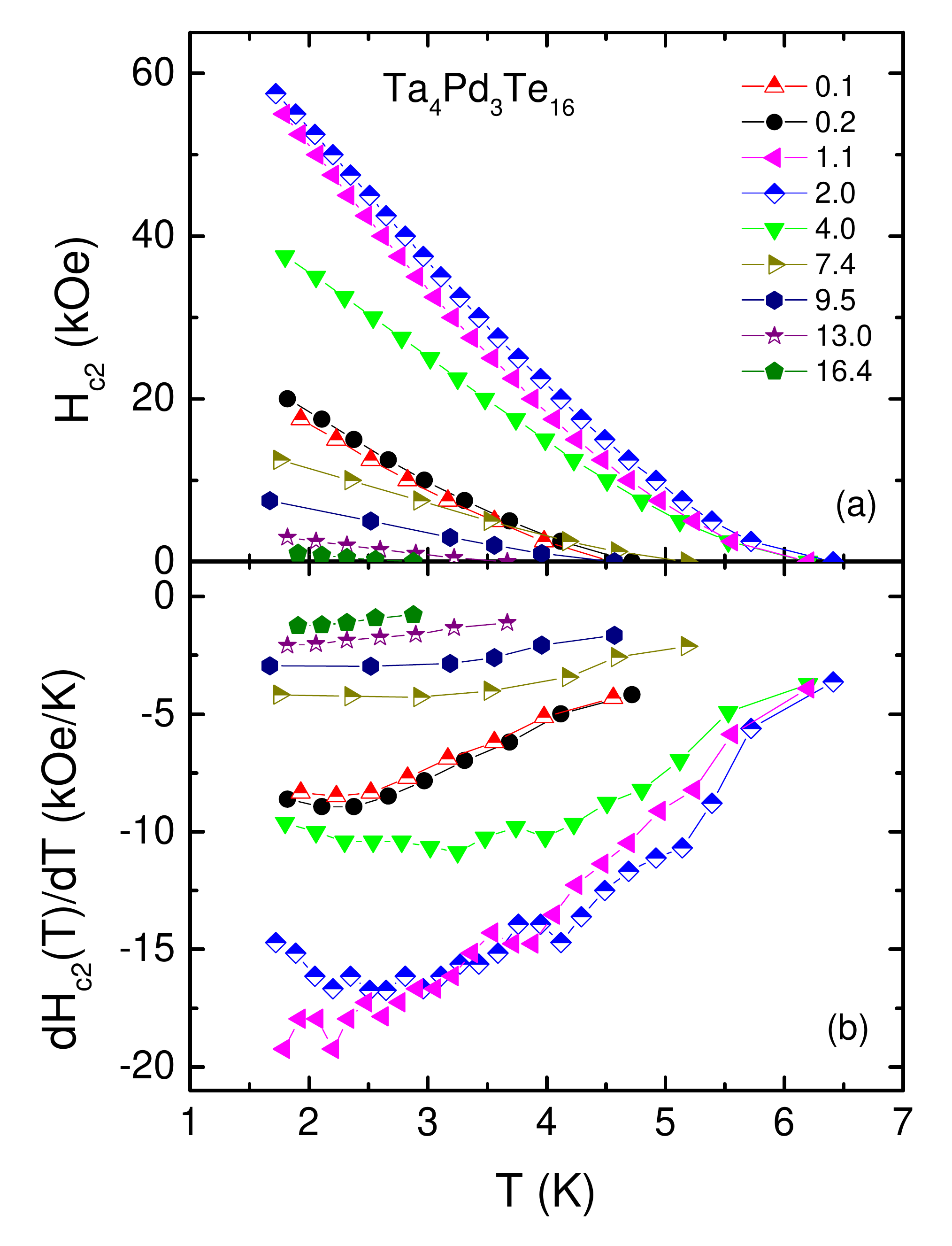}
\end{center}
\caption{(Color online) (a) Temperature dependent upper critical field (for $H \| (-1~0~3)$) measured at different pressures (pressure values are given in kbar). (b) Corresponding derivatives, $dH_{c2}/dT$, at different pressures.} \label{F6}
\end{figure}

\begin{figure}
\begin{center}
\includegraphics[angle=0,width=120mm]{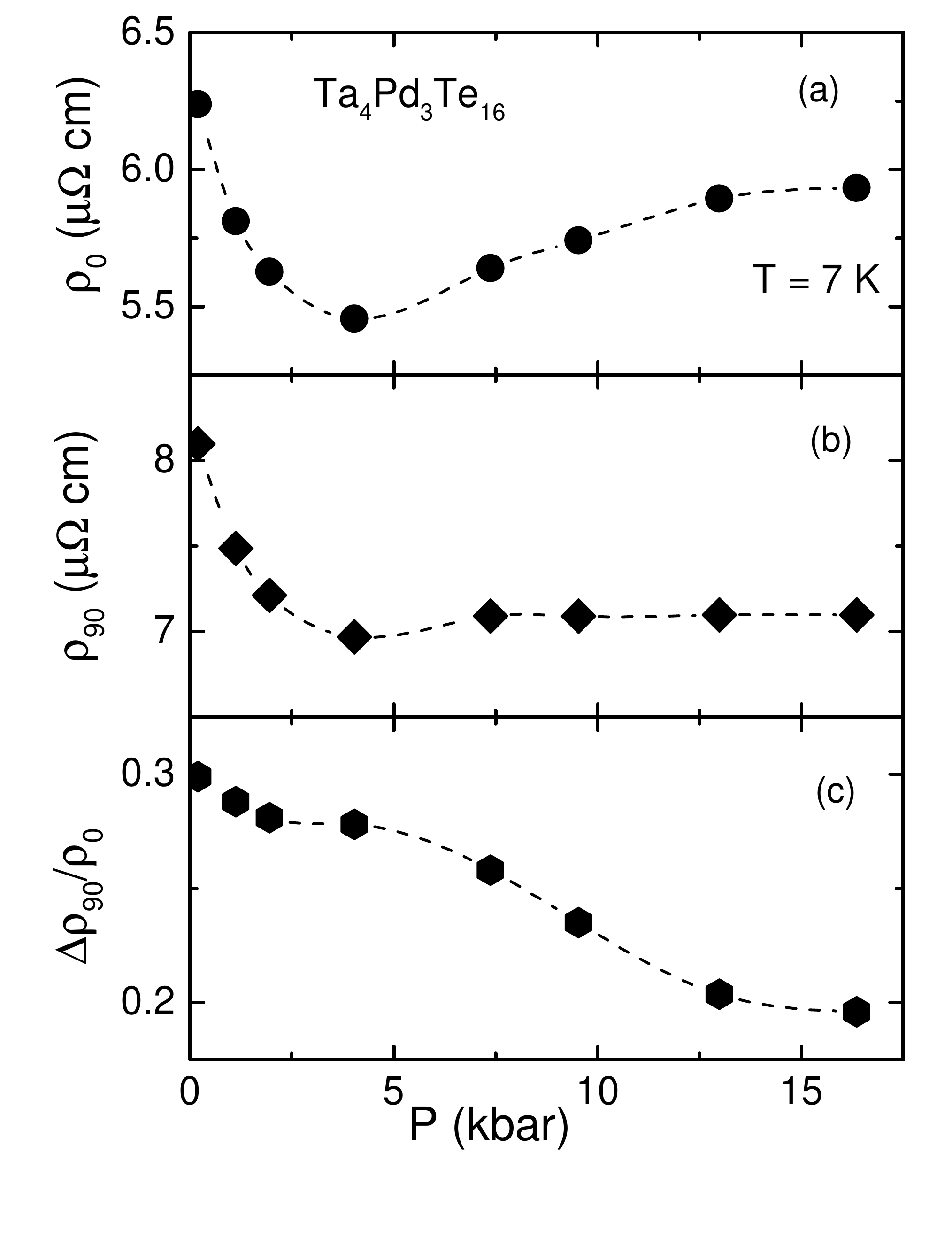}
\end{center}
\caption{ Pressure dependence of (a) zero field resistivity, $\rho_0$; (b) resistivity in $H = 90$ kOe magnetic field, $\rho_{90}$; and (c) magnetoresistivity, $\Delta\rho_{90}/\rho_0 = (\rho_{H = 90~\text{kOe}} - \rho_{H = 0})/\rho_{H = 0}$, measured at $T = 7$ K. Dashed lines are guides for the eye.} \label{F11}
\end{figure}

As presented above, all measured in this work superconducting ($T_c, H_{c2}$) and normal state (zero field resistivity, normal state magnetoresistivity) properties have anomalies in the 2 - 4 kbar pressure range. Although these data, given elusive signatures in bulk measurements and our experimental restrictions,  provide no direct evidence of the existence of CDW in \TPT~ either at ambient or at high pressure, they are consistent with the hypothesis of coexistence of CDW and superconductivity at ambient pressure.  Within the same hypothesis this  CDW is suppressed either to $T = 0$ K or, at least below the superconducting transition temperature, at 2 - 4 kbar. \cite{pan15a}  At low pressures $T_c$ increases under pressure with the initial slope, $dT_c/dP|_{P = 0} \approx 1.5$ K/kbar (Fig. \ref{F4}) and normal state, low temperature resistivity decreases (Fig. \ref{F11}(a)). This $T_c$ behavior appears to be consistent with Friedel's picture of increase of electron density by closing the gaps at the Fermi surface as the CDW state is suppressed. \cite{fri75a,jer76a} Another, indirect suggestion of coexistence of superconductivity and CDW at low pressures is significant positive curvature of $H_{c2}(T)$ \cite{mac81a,gab88a} that becomes less pronounced above $\sim 4$ kbar (Fig. \ref{F6}).

At pressures above 4 kbar the $T_c$, $H_{c2}(0)$, and the normalized temperature derivative of $H_{c2}$, $[-dH_{c2}/dT]/T_c$, decrease with increase of pressure, whereas the low temperature normal state resistivity, as well as the Debye temperature estimated from the Bloch - Gr\"uneisen - Mott fits of resistivity both increase. In a simple case of an anisotropic superconductor, in a clean limit, \cite{kog12a}  $[-dH_{c2}/dT]/T_c \propto 1/v_F^2$ , where $v_F$ is the Fermi velocity. Within this simple model the experimentally observed  $[-dH_{c2}/dT]/T_c$  behavior can be accounted for if the $v_F^2$ increases under pressure.  Electrical resistivity in an isotropic model with elastic electron scattering can be written as $\rho \propto 1/(v_F^2 \tau D_F)$, \cite{abr88a} where $\tau$ is the scattering time and $D_F$ is the density of states at the Fermi level. Then, for consistent description of  $[-dH_{c2}/dT]/T_c$ and $\rho$, the density of states at the Fermi level should decrease under pressure faster than $v_F^2$ increases (we assume that the scattering time $\tau$ is pressure independent).

\begin{figure}
\begin{center}
\includegraphics[angle=0,width=120mm]{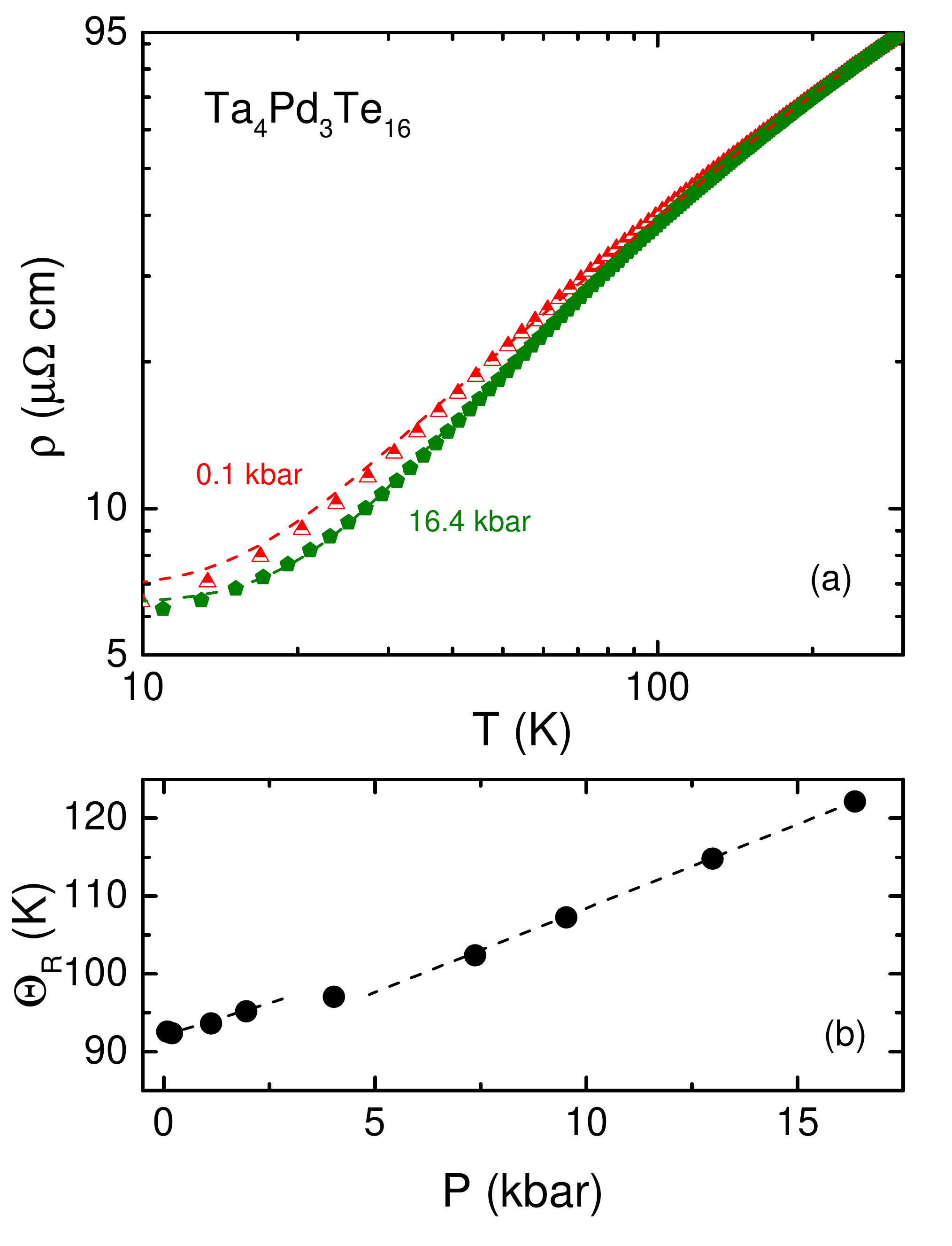}
\end{center}
\caption{(Color online) (a) Data (symbols) and Bloch - Gr\"uneisen - Mott fits (see the text) for $\rho(T)$ data taken at 0.1 and 16.4 kbar (b) pressure dependence of the Debye temperature, $\Theta_R$, obtained from the fits. Dashed lines are guides for the eye. The error bars for the values of $\Theta_R$ obtained as a result of fits are $\leq 1.1$ K, i.e.  smaller than the size of the symbols. } \label{F12}
\end{figure}

Now we can turn to negative $dT_c/dP$ above $\sim 4$ kbar. If we ignore possible changes under pressure in the Coulomb pseudopotential and efffective electron-phonon interaction \cite{lor05a}, the increase of the Debye temperature under pressure alone (Fig. \ref{F12}(b)) would cause an increase of $T_c$, however the decrease of density of states with pressure evidently dominates, resulting in the $T_c$ decrease.

All in all, the experimental observations over the whole studied pressure range can be qualitatively understood by considering the density of states at the Fermi level a dominant parameter.
By application of pressure, Fermi level passes through a shap maximum in the density of states. Generally speaking this can be realized without CDW, by having at ambient pressure a flat, pressure - sensitive band close to the Fermi level. However, given experimental data that suggest existence of CDW at ambient pressure \cite{fan15a,duz15a,liz16a,hel17a} it is possible that the following scenario is realized.  At low pressures $D_F$ initially increases due to  closing of the gaps at the Fermi surface as the CDW state is suppressed. At higher pressures, after CDW is suppressed, $D_F$ decreases with pressure. 

Further measurements under pressure as well as band structural calculations under pressure would be desirable to directly assess the $D_F(P)$ behavior. X-ray scattering measurements at ambient and elevated pressures would be desirable to to understand the nature of the suggested CDW state.

\begin{acknowledgments}
We appreciate fecund discussions with Vladimir Kogan. Research was supported by the U.S. Department of Energy, Office of Basic Energy Sciences, Division of Materials Sciences and Engineering. Ames Laboratory is operated for the U.S. Department of Energy by the Iowa State University under Contract No. DE-AC02-07CH11358. Na Hyun Jo was supported by the Gordon and Betty Moore Foundation EPiQS Initiative (Grant No. GBMF4411).
\end{acknowledgments}

\clearpage

\clearpage

\clearpage

\clearpage

\clearpage

\clearpage

\clearpage

\clearpage

\clearpage

\clearpage

\clearpage

\begin{thebibliography}{99}

\bibitem{mar91a}
Arthur Mar and James A. Ibers, J. Chem. Soc. Dalton Trans. {\bf S} 693 (1991).

\bibitem{ale97a}
Pere Alemany, St\'ephane Jobic, Raymond Brec, and Enric Canadell, Inorg. Chem. {\bf 36}, 5050 (1997).

\bibitem{sin14a}
David J. Singh, Phys. Rev. B {\bf 90}, 144501 (2014).

\bibitem{lee15a}
Wang-Ro Lee, Sung-Woo Cho, Dae-Hyun Nam, and Dongwoon Jung, Bull. Korean Chem. Soc. {\bf 36}, 1859 (2015).

\bibitem{jia14a}
Wen-He Jiao, Zhang-Tu Tang, Yun-Lei Sun, Yi Liu, Qian Tao, Chun-Mu Feng, Yue-Wu Zeng, ZhAn Xu, and Guang-Han Cao, J. Am. Chem. Soc. {\bf 136}, 1284 (2014).

\bibitem{pan15a}
J. Pan, W. H. Jiao, X. C. Hong, Z. Zhang, L. P. He, P. L. Cai, J. Zhang, G. H. Cao, and S. Y. Li, Phys. Rev. B {\bf 92}, 180505 (2015).

\bibitem{jia15a}
Wen-He Jiao, Yi Liu, Yu-Ke Li, Xiao-Feng Xu, Jin-Ke Bao, Chun-Mu Feng, S. Y. Li, Zhu-An Xu and Guang-Han Cao, J. Phys.: Condens. Matter {\bf 27} 325701 (2015).

\bibitem{xux15a}
Xiaofeng Xu, W. H. Jiao, N. Zhou, Y. Guo, Y. K. Li, Jianhui Dai, Z. Q. Lin, Y. J. Liu, Zengwei Zhu, Xin Lu, H. Q. Yuan, and Guanghan Cao, J. Phys.: Condens. Matter {\bf 27} 335701 (2015).

\bibitem{zha16a}
Q. R. Zhang, D. Rhodes, B. Zeng, M. D. Johannes, and L. Balicas, Phys. Rev. B {\bf 94}, 094511 (2016).

\bibitem{fan15a}
Q. Fan, W. H. Zhang, X. Liu, Y. J. Yan, M. Q. Ren, M. Xia, H. Y. Chen, D. F. Xu, Z. R. Ye, W. H. Jiao, G. H. Cao, B. P. Xie, T. Zhang, and D. L. Feng, Phys. Rev. B {\bf 91}, 104506 (2015).

\bibitem{duz15a}
Zengyi Du, Delong Fang, Zhenyu Wang, Yufeng Li, Guan Du, Huan Yang, Xiyu Zhu, and  Hai-Hu Wen, Sci. Rep. {\bf 5}, 9408 (2015).

\bibitem{liz16a}
Z. Li, W. H. Jiao, G. H. Cao, and Guo-qing Zheng, Phys. Rev. B {\bf 94}, 174511 (2016).

\bibitem{che15a}
D. Chen, P. Richard, Z.-D. Song, W.-L. Zhang, S.-F. Wu, W. H. Jiao, Z. Fang, G.-H. Cao, and H. Ding, J. Phys.: Condens. Matter {\bf 27}, 495701 (2015).

\bibitem{hel17a}
T. Helm, F. Flicker, R. Kealhofer, P. J. W. Moll, I. M. Hayes, N. P. Breznay, Z. Li, S. G. Louie, Q. R. Zhang, L. Balicas, J. E. Moore, J. G. Analytis, Phys. Rev. B {\bf 95}, 075121 (2017).

\bibitem{gab02a}
A. M. Gabovich, A. I. Voitenko, and M. Ausloos, Phys. Rep. {\bf 367}, 583 (2002).

\bibitem{tau14a}
Valentin Taufour, Neda Foroozani, Makariy A. Tanatar, Jinhyuk Lim, Udhara Kaluarachchi, Stella K. Kim, Yong Liu, Thomas A. Lograsso, Vladimir G. Kogan, Ruslan Prozorov, Sergey L. Bud'ko, James S. Schilling, and Paul C. Canfield, Phys. Rev. B {\bf 89}, 220509 (2014).

\bibitem{kal16a}
Udhara S. Kaluarachchi, Valentin Taufour, Anna E. Böhmer, Makariy A. Tanatar, Sergey L. Bud'ko, Vladimir G. Kogan, Ruslan Prozorov, and Paul C. Canfield, Phys. Rev. B {\bf 93}, 064503 (2016) 

\bibitem{can92a}
P. C. Canfield and Z. Fisk, Philos. Mag. B {\bf 65}, 1117 (1992).

\bibitem{can16a}
 P. C. Canfield, T. Kong, U. S. Kaluarachchi, and N. H. Jo, Philos. Mag. {\bf 96}, 84 (2016).

\bibitem{jes16a}
A. Jesche, M. Fix, A. Kreyssig, W. R. Meier, and P. C. Canfield, Philos. Mag. {\bf 96}, 2115 (2016).

\bibitem{jia16a}
 W.-H. Jiao, L.-P. He, Y. Liu, X.-F. Xu, Y.-K. Li, C.-H.Zhang, N. Zhou, Z.-A. Xu, S.-Y. Li, and G.-H. Cao, Scientific Reports {\bf 6}, 21628 (2016).

\bibitem{HMD}
\url{https://www.qdusa.com/sitedocs/productBrochures/High_Pressure_Cell_for_Magnetometry_Brochure.pdf}

\bibitem{yok07a}
Keiichi Yokogawa, Keizo Murata, Harukazu Yoshino, and Shoji Aoyama, Jpn. J. Appl. Phys. {\bf 46}, 3636 (2007).

\bibitem{eil81a}
A. Eiling and J. S. Schilling, J. Phys. F: Metal Phys. {\bf 11}, 623 (1981).

\bibitem{bud84a}
 S. L. Bud'ko, A. N. Voronovskii, A. G. Gapotchenko, and E. S. Itskevich, Zh. Eksp. Teor. Fiz. {\bf 86}, 778 (1984), [Sov. Phys. JETP {\bf 59}, 454 (1984)].

\bibitem{tor15a}
 M. S. Torikachvili, S. K. Kim, E. Colombier, S. L. Bud'ko, and P. C. Canfield, Rev. Sci. Instrum. {\bf 86}, 123904 (2015).

\bibitem{kac00a}
D. Kaczorowski, B. Andraka, R. Pietri, T. Cichorek, and V. I. Zaremba, Phys. Rev. B {\bf 61}, 15255 (2000).

\bibitem{fri75a}
J. Friedel, J. Physique - Lett. {\bf 36}, L-279 (1975).

\bibitem{jer76a}
D. J\'erome, C. Berthier, P. Molini\'e, and J. Rouxel, J. Physique, Colloque {\bf 4}, 125 (1976).

\bibitem{mac81a}
Kazushige Machida, Tamotsu K\=oyama, and Takeo Matsubara, Phys. Rev. B {\bf 23}, 99 (1981).

\bibitem{gab88a}
A. M. Gabovich and A. S. Shpigel, Phys. Rev. B {\bf 38}, 297 (1988).

\bibitem{kog12a}
V. G. Kogan and R. Prozorov, Rep. Prog. Phys. {\bf 75}, 114502 (2012).

\bibitem{abr88a}
A. A. Abrikosov, Fundamentals of the theory of metals, North-Holland, Amsterdam, 1988.

\bibitem{lor05a}
B. Lorentz and C. W. Chu, in: Frontiers in Superconducting Materials, ed. by A. V. Narlikar (Springer, Berlin, 2005)  p. 459.



\end{thebibliography}
\end{document}